\documentclass[12pt]{article} 
\usepackage{amssymb,amsmath,color}

\numberwithin{equation}{section}
\usepackage[toc,page]{appendix}

\newcommand{\nc}{\newcommand}

\nc{\tp}{{\cal P}}
\nc{\la}{\lambda} \nc{\alf}{\alpha} \nc{\La}{\Lambda} \nc{\ze}{\zeta}
\nc{\tht}{\theta} \nc{\T}{\Theta} \nc{\be}{\beta}  \nc{\eps}{\epsilon} 
\nc{\ga}{\gamma}  \nc{\De}{\Delta}  \nc{\G}{\Gamma}  \nc{\vphi}{\varphi}
\nc{\de}{\delta} \nc{\si}{\sigma}  \nc{\ka}{\kappa}   \nc{\Si}{\Sigma} 
\nc{\om}{\omega}  \nc{\qq}{\quad\quad}                \nc{\Om}{\Omega}
\nc{\nf}{\infty}   \nc{\dl}{\mathop{\smash{\cal L}}}  \nc{\black}{\rule{3mm}{3mm}}
\nc{\ra}{\rightarrow}    \nc{\ol}{\overline}        \nc{\und}{\underline} 
\nc{\beq}{\begin{equation}}  \nc{\eeq}{\end{equation}}  \nc{\pt}{\partial}  
   \nc{\dst}{\displaystyle}  \nc{\na}{\nabla} 
\nc{\nnb}{\nonumber}    \nc{\bs}{\backslash}        \nc{\mb}{\mathbb}   
\nc{\sn}{{\rm sn}\,} \nc{\cn}{{\rm cn}\,}     \nc{\dn}{{\rm dn}\,} \nc{\nin}{\noindent}
\nc{\ti}{\tilde}   \nc{\wti}{\widetilde}   \nc{\h}{\hat}  \nc{\wh}{\widehat}
\nc{\tpsi}{\wti{\psi}}   \nc{\tphi}{\wti{\phi}}  \nc{\tH}{\wti{H}} \nc{\Ai}{{\rm Ai}}

\newcounter{muni}
\newenvironment{remunerate}{\begin{list}{{\rm \arabic{muni}.}}
{\usecounter{muni}
\setlength{\leftmargin}{0pt}\setlength{\itemindent}{38pt}}}{\end{list}}

\nc{\brm}{\begin{remunerate}}   \nc{\erm}{\end{remunerate}}

\newtheorem{nth}{Proposition}  \newtheorem{nTh}{Theorem}
\nc{\stg}{\mathop{\smash{*}}}
\nc{\st}{\mathop{\smash{\delta}}}
\nc{\barr}{\begin{array}}   \nc{\earr}{\end{array}}   \nc{\dg}{\dagger}
\nc{\mtvb}{\mathversion{bold}}   \nc{\mtvn}{\mathversion{normal}}  \nc{\F}{f_{\eps}}

\begin{document} 

\begin{titlepage}

\date{\today}

\vspace{1cm}
\centerline{\huge\bf  Superintegrable models}

\vspace{5mm}
\centerline{\huge\bf on riemannian surfaces of revolution}

\vspace{5mm} 
\centerline{\huge\bf with integrals of any integer degree (II)}

\vspace{2cm}
\centerline{\large\bf  Galliano VALENT }

\vspace{1cm}
\centerline{ \it Laboratoire de Physique Math\'ematique de Provence}
\centerline{\it Avenue Marius Jouveau 1, 13090 Aix-en-Provence, France}

\vspace{2cm}

\begin{abstract}The construction of Superintegrable models with rotational symmetry and two integrals of any integer degree greater than 3 was completed in \cite{Va1} only for the so called {\em simple} case. It is extended here to a more general situation and several globally defined examples are worked out, all of them living either in ${\mb R}^2$ or in ${\mb H}^2$.
\end{abstract}

\vspace{3cm}
MSC 2010 numbers: {\tt 32C05}, {\tt 81V99}, {\tt 37E99}, {\tt  37K25}.

\end{titlepage}

\tableofcontents

\section{Introduction}
Matveev and Shevchishin in \cite{ms} have introduced a new area of research in the field of superintegrable (SI) two dimensional models. Their starting point is the metric 
\beq
g=\frac{dt^2+dy^2}{h_t^2}\qq\qq h_t=\frac{dh(t)}{dt}
\eeq
which describes a surface of revolution. It follows that the geodesic flow, having for hamiltonian
\beq
H=h_t^2(P_t^2+P_y^2)\eeq
has a linear first integral $P_y$ and is therefore integrable in Liouville sense.

Matveev and Shevchishin  considered, in the phase space, the linear mapping ${\cal L}:\ F\,\to\,\{P_y,F\} $ and classified  the extra integrals (leading to SI) according to the eigenvalues $\mu$ of ${\cal L}$. This leads to three cases 
\beq\label{3cas}\barr{cll}
\mu=\pm i & \qq \mbox{trigonometric case} \qq & S_2=\sin y\,A+\cos y\,B,\\[4mm]
\mu=\pm 1 & \qq \mbox{hyperbolic case} \qq & S_2=\sinh y \,A+\cosh y\,B,\\[4mm]
\mu=0 & \qq \mbox{affine case} \qq & \dst S_2=A+y\,B+\frac{y^2}{2}C,
\earr\eeq
where all the functions depend on $(t,\,P_t,\,P_y)$. Due to Poisson theorem it follows that $S_1=\{P_y,S_2\}$ is also an integral, leading, in all of the three cases, to 3 integrals besides $H$, which cannot be algebraically independent.

In the article \cite{ms} the case of coefficients 
$(A,B,C)$ quadratic in the momenta was shown to lead back to Koenigs metrics \cite{Ko}, generalized in \cite{kkmw}. The quadratic integrals led in \cite{Va2} to a simple determination of the geodesics. However this class of metrics never meets the  manifold ${\mb S}^2$ and this induced Matveev and Shevchishin to try with cubic integrals. They succeeded to reduce the construction of the metric to a non-linear, first order ODE for $h(t)$ in all of the three cases considered above.

These ODEs were solved in \cite{vds} for all of the three cases, but only the trigonometric case led to a two parameter metric defined on ${\mb S}^2$ which was proved in \cite{VaLMP} to be of Zoll type (all of its geodesics have for length $2\pi$). Even better there appear also new metrics, not of Zoll type, but with geodesics which are all closed, defined on Tannery's teardrop orbifold, possibly the first appearance of SI models defined on orbifolds.

It was realized that, in the three classes appearing in (\ref{3cas}), the affine case could be simpler and indeed in \cite{Va1} the explicit form of the metric was obtained for extra integrals of any integer degree larger or equal to 3. 

The core of this construction is that, given a monic polynomial of degree $n$
\beq
F(a)=a^n+\sum_{k=0}^{n-1}\,A_k\,a^k,\eeq
one has to solve a nth order {\em linear} and  {\em homogeneous} ODE for the function $x(a)$, which is 
\beq
\sum_{s=0}^n\frac{F^{(n-s)}}{(n-s)!}\frac{D_a^s\,x}{(1/2)_s}=0
\eeq
in the case of integrals of degree $2n$, and a nth order {\em linear} but no longer homogeneous ODE \footnote{The constant $\be_n$ appearing in \cite{Va1} can be omitted since only $x'(a)$ appears in the metric.}
\beq
\sum_{s=0}^n\frac{F^{(n-s)}}{(n-s)!}\frac{D_a^s\,x}{(1/2)_s}=\Big(n+\frac 12\Big)\nu_n\,a
\eeq
in the case of integrals of degree $2n+1$.
 
In \cite{Va1}, for a {\em simple} $F$ (for which all of its roots are different), we determined explicitly the function $x$ and obtained an explicit construction of the metric and of the integrals $S_1$ and $S_2$. 
   
It is the aim of this article to generalize these results    to the case where one (or several) real roots are multiple. This seemingly easy generalization requires as much work as the {\em simple}  case.

The content of this article is the following: in Section 2 we consider the case of integrals of even degrees in the momenta. After a bird's eye view of the {\em simple} case, the generalization to a multiple real root is done and the integrals determined. For quartic integrals we establish the relation with Novichkov results. Two globally defined examples are worked out.

The structure of Section 3 is similar but covers the case of odd degree integrals. Let us notice that in all cases we meet only non-compact manifolds, namely ${\mb R}^2$ and ${\mb H}^2$.

Some open problems are discussed in the concluding Section 4.

\section{Integrals of even degrees in the momenta}
Let us first recall the results obtained in \cite{Va1}.

\subsection{The simple case}

\subsubsection{General setting}
Let us define the objects
\beq
F(a)=a^n+\sum_{k=0}^{n-1}\,A_k\,a^k \qq\qq G=H^n+\sum_{k=0}^{n-1}\,A_k\,H^k\,P_y^{2(n-k)},\eeq
where the string $\ (A_0,\,A_1,\ldots,A_{n-1})\ $ of free real constants does parametrize our construction. 

The metric and the hamiltonian,  in the phase space of coordinates $\ (a,y,P_a,P_y)\ $, are
\beq
g=\left(\frac{x'}{a}\right)^2 da^2+\frac{dy^2}{a}\qq\qq H=\Pi^2+a\,P_y^2\qq \Pi=\frac a{x'}\,P_a.
\eeq 
This system is defined on a surface of revolution because  $x$ depends solely on the coordinate $a$, implying 
$\,\{H,P_y\}=0$ and therefore integrability in Liouville sense.

\subsubsection{Extra integrals}
This dynamical system will lift up to a SI one if we can  find one extra integral given by: \footnote{Notice that with respect to \cite{Va1} the notations $\wti{b}_k$ and $b_k$ have been exchanged.}
\beq
S_1=Q_1+yG\qq\qq  Q_1=\sum_{k=1}^n\,\wti{b}_k[F]\,\Pi^{2k-1}\,P_y^{2(n-k)+1}\qq\quad\sharp[S_1]=\sharp[Q_1]=2n.
\eeq 
As shown in \cite{Va1}, if $x(a)$ is a solution of the nth order linear and homogeneous ODE \footnote{The Pochammer symbols are defined by $\dst (a)_n=\frac{\G(a+n)}{\G(a)}$.}
\beq
Op_n[F]\,x\equiv\sum_{s=0}^n\,\frac{F^{(n-s)}}{(n-s)!}\,\frac{D_a^s\,x}{(1/2)_s}=0\eeq
and if 
\beq
\wti{b}_k[F]=\sum_{s=1}^k\,\frac{F^{(k-s)}}{(k-s)!}\,\frac{D_a^s\,x}{(1/2)_s}\eeq
then $S_1$ is indeed an integral of $H$.

In practice it is more convenient to define new functions $b_k$, allowing to write
\beq
Q_1=\sum_{k=1}^n\,b_k[F]\,H^{n-k}\,\Pi\,P_y^{2k-1},
\eeq
and given by
\beq
b_k[F](a)=\sum_{s=1}^k\,{n-s \choose k-s}(-a)^{k-s}
\,\wti{b}_{n-s+1}(a).\eeq

In fact, a second integral of degree $2n$ can be exhibited
\beq
S_2=Q_2+y\,Q_1+\frac{y^2}{2}\,G \qq\qq Q_2=\sum_{k=1}^n\,c_k[F]\,H^{n-k}\,P_y^{2k}\qq\quad \sharp[Q_2]=2n\eeq
provided that
\beq
c'_k=-b_k\,x' \qq\qq k\in\{1,2,\ldots,n\}.
\eeq

In \cite{Va1} the {\em simple} case, where all the roots of $F$ are simple (with symbol $\wh{F}$), was completely solved. The function $x$ is given by
\beq\label{xpair}
x(a)=\sum_{i=1}^n\,\frac{\xi_i}{\sqrt{\De_i}}\qq\qq \De_i=\eps_i\,(a-a_i)\qq \eps_i^2=1,
\eeq
the integral $S_1$ (resp. $S_2$) involves the functions
\beq\label{bcpair}\left\{\barr{l}\dst 
 b_k[\wh{F}]=(-1)^k\,\sum_{i=1}^n\,\frac{\xi_i}{\sqrt{\De_i}}\,\si^i_{k-1},\\[4mm]\dst 
c_k[\wh{F}]=\frac{(-1)^{k+1}}{2}\left(\sum_{i=1}^n\frac{\xi_i^2}{\De_i}\,\si^i_{k-1}+
\sum_{i\neq j=1}^n\,\frac{\xi_i\,\xi_j}{\sqrt{\De_i\,\De_j}}\Big(\si^{ij}_{k-1}+a\,\si^{ij}_{k-2}\Big)\right),\earr\right.\eeq
for $k\in\,\{1,2,\ldots n\}$.

\subsubsection{Relation with Novichkov results}
In \cite{ms} it was shown that the construction of a SI sytem with one linear integral and an extra cubic one could be reduced to a highly non-linear first order differential equation. This approach was generalized by Novichkov in his Thesis \cite{No} to the case of an extra quartic integral i. e. where $n=2$ so that $F(a)=(a-a_1)(a-a_2)$. 

Let us first summarize his results. To avoid clutter we will use for him the coordinates 
$(t,y,P_t,P_y)$ and will transform his constants $A_i$ into $\,B_i$. His metric and hamiltonian are
\beq
g=\frac{dt^2+dy^2}{h_t^2}\qq h_t=D_t\,h(t) \qq\qq H=\Pi^2+h_t^2\,P_y^2\qq \Pi=h_t\,P_t
\eeq
and he considers the quartic integral $F=Q_1+yG$ with
\beq
Q_1=\Big[(B_1-B_0\,h)h_t^2-\frac{(B_3+B_4\,t)}{h_t}\Big]\,\Pi\,P_y^3+(B_1-B_0\,h)\,\Pi^3\,P_y
\eeq
and
\beq 
G=-B_4\,P_y^4+B_2\,H\,P_y^2+B_0\,H^2.
\eeq
One of the results of his Thesis is the following:

\begin{nTh} The model considered is SI either if we have
\beq\label{N1}
2h_t^2\,{\cal H}{\cal T}-B_4\,{\cal H}^2+\left(\frac{B_2}{B_4}h_t^2-1\right){\cal T}^2=B_5\eeq
or if we have
\beq\label{N2}
(h_t^2+B_2)\,{\cal H}^2+\frac{h_t^2}{B_4}{\cal T}^2-2{\cal H}{\cal T}=B_6\eeq
for some constants $(B_5,B_6)$ where
\beq
{\cal H}(t)=h(t)-B_1\qq\qq {\cal T}(t)=\frac{B_3+B_4\,t}{h_t}.\eeq
\end{nTh} 
Let us explain how we solve Novichkov equations: 

\begin{nth} For $a_1\neq a_2$, Novichkov equations are solved by the following parametric representation:
\beq\label{N3}\left\{\barr{l}\dst 
{\cal H}(t)=x(a)=\frac{\xi_1}{\sqrt{a-a_1}}+\frac{\xi_2}{\sqrt{a-a_2}}\\[4mm]\dst {\cal T}(t)=-\left(\frac{a_2\,\xi_1}{\sqrt{a-a_1}}+\frac{a_1\,\xi_2}{\sqrt{a-a_2}}\right)\earr\right. \qq\quad h_t=\sqrt{a}.
\eeq
\end{nth}

\nin{\bf Proof:} Comparing the function $G$ in both approaches gives the relations
\beq
B_0=1\qq\qq B_2=A_1=-(a_1+a_2)\qq\qq B_4=-A_0=-a_1\,a_2.
\eeq
Comparing the metrics gives the relations
\beq\label{Nov1}
h_t=\sqrt{a}\qq\qq dt=\frac{x'}{\sqrt{a}}\,da\quad\Longrightarrow\quad dh=x'\,da \quad\Longrightarrow\quad {\cal H}(t)=x(a),\eeq
where
\beq
x(a)=\frac{\xi_1}{\sqrt{a-a_1}}+\frac{\xi_2}{\sqrt{a-a_2}}\qq\qq a_1\neq a_2.\eeq
Integrating for $t$ we get
\beq\label{Nov2}
t=t_0+\sqrt{a}\left(\frac{\xi_1}{a_1\,\sqrt{a-a_1}}+\frac{\xi_2}{a_2\,\sqrt{a-a_2}}\right).\eeq
Comparing $Q_1$ in each approach gives $\ {\cal T}(t)=-(a\,x+2Fx')\ $ which has to 
match with (\ref{Nov2}) giving $A_0\,t_0=B_3$ and leaving us with the formula for $\,{\cal T}(t)\,$ given in   (\ref{N3}).

Novichkov equations (\ref{N1}) and (\ref{N2}) become:
\beq\left\{\barr{l}\dst
B_5=2a\,{\cal H}{\cal T}+A_0\,{\cal H}^2-\frac{A_1}{A_0}a\,{\cal T}^2-{\cal T}^2,\\[4mm]\dst
B_6=a\,{\cal H}^2+A_1\,{\cal H}^2-\frac{a}{A_0}{\cal T}^2-2{\cal H}{\cal T}.\earr\right.
\eeq
Checking these relations gives for the constants
\beq
B_5=\frac{(a_1-a_2)}{a_1a_2}(a_1^2\xi_2^2-a_2^2\xi_1^2)\qq\qq 
B_6=-\frac{(a_1-a_2)}{a_1a_2}(a_1\xi_2^2-a_2\xi_1^2),
\eeq
and ends up the proof.$\hfill\Box$

Now let us turn ourselves to the case of a {\em multiple real root} for $F$.

\subsection{The case of a multiple real root}
Let it be supposed that a single real root of $F$, let us say $a_1$, has multiplicity $r$ while all other roots remain simple: 
\beq
r\in\,\{2,\ldots,n\}: \qq F(a)=(a-a_1)^r\,\wh{F}\qq\qq \wh{F}=\prod_{i=r+1}^n\,(a-a_i)
\eeq

\subsubsection{Determining ${\mathbf x}$}
Let us begin with:

\begin{nth}\label{prop2} The function $x$ is given by 
\beq
x(a)=\sum_{l=1}^r\,\frac{\mu_l}{\De_1^{l-1/2}}+\sum_{i=r+1}^n\,\frac{\xi_i}{\sqrt{\De_i}}\eeq
Notice that if $r=n$ one has to take $\wh{F}=1$ and the sum over the $\xi_i$ as vanishing.
\end{nth}

\nin{\bf Proof:} 

Due to the linearity of the ODE for $x$ we just need to prove that
\beq
Op_n[F]\,\De_1^{-l+1/2}=0\qq\qq l\in\,\{1,2,\ldots,r\}.
\eeq
Using Appendix A we have
\beq
Op_n[F]\,\De_1^{-l+1/2}=B_{0,n,l}[F]={\cal P}_l\ \sum_{s=1}^l\frac{(-1/2)_{l-s}}{(l-s)!}\ \De_1^{s-l+1/2}\ \frac{D_a^{n}}{n!}\left(\frac F{\De_1^s}\right).
\eeq
Observing that the polynomial $F/\De_1^s$ has a degree 
$\leq n-1$ concludes the proof. $\hfill\Box$

Let us consider now the integral $S_1$.

\subsubsection{Determining ${\mathbf S_1}$}
We have seen that $\ S_1=Q_1+y\,G$ where
\beq\label{defQ1}
Q_1=\sum_{k=1}^n\,b_k[F]\,H^{n-k}\,\Pi\,P_y^{2k-1}.
\eeq
Let us prove 
\begin{nth}\label{Q1} Starting from\qq\qq 
\beq
x(a)=\sum_{l=1}^r\ \frac{\mu_l}{\De_1^{l-1/2}}+\sum_{i=r+1}^n\frac{\xi_i}{\sqrt{\De_i}},\qq\qq \tp_l=\frac{(1/2)_{l-1}}{(l-1)!},
\eeq
the functions $b_k$, for $k\in\{1,2,\ldots,n\}$, are given by \footnote{As usual, for $r=n$, the sum over $\xi_i$ should disappear.}
\beq\label{formQ1}
b_k[F]=(-1)^k\left(\sum_{l=1}^r\,\frac{\mu_l}{\tp_l}\,\sum_{s=1}^{l}(-\eps_1)^{s-1}\,\si^{(1,s)}_{k-s}\,
\frac{\tp_{l-s+1}}{\De_1^{l-s+1/2}}
+\sum_{i=r+1}^n\frac{\xi_i}{\sqrt{\De_i}}\,\si^i_{k-1}\right).
\eeq
where the $\si^{(1,s)}_{k-s}$ are defined in the 
Appendix B.
\end{nth}

\nin{\bf Proof:} As we have seen the function $x$ is made out of two pieces: the first one linear in the parameters $\mu_l$ and the second one linear on the parameters 
$\xi_i$. We will not compute the contribution to 
$\wti{b}_k$ of these last terms because they are the 
same as for the {\em simple} case. We will indicate 
this fact by using the symbol $\cong$ instead of the symbol $=$.

Let us start from
\beq
\wti{b}_k[F]\cong\sum_{l=1}^r\,\mu_l\sum_{s=1}^k\frac{F^{(k-s)}}{(k-s)!}\,\frac{D_a^s\,x_l}{(1/2)_s}=\sum_{s=1}^k\,\mu_l\,B_{1,k,l},\qq x_l=\frac 1{\De_1^{l-1/2}},
\eeq
where $B_{1,k,l}$ is computed in Appendix A. We get 
\beq
\wti{b}_k[F]\cong(-\eps_1)\sum_{l=1}^r\,\frac{\mu_l}{\tp_l}\,\sum_{s=1}^l\,\frac{\tp_{l-s+1}}{\De_1^{l-s+1/2}}\,\frac{D_a^{k-1}}{(k-1)!}\left(\frac F{\De_1^s}\right),
\eeq
from which $\wti{b}_{n-t+1}$ follows. 

Using relation (\ref{bdef1}) we have
\beq
\frac F{\De_1^s}=\eps_1^s\,\sum_{\nu=0}^{n+s}(-1)^{\nu-s}\,\si^{(1,\,s)}_{\nu-s}\,a^{n-\nu},
\eeq
which entails
\beq
\wti{b}_{n-t+1}[F]\cong\sum_{l=1}^r\frac{\mu_l}{\tp_l}\sum_{s=1}^l\,(-\eps_1)^{s-1}\frac{\tp_{l-s+1}}{\De_1^{l-s+1/2}}\sum_{\nu=0}^{n+s}(-1)^{\nu}\,\si^{(1,\,s)}_{\nu-s}\,\frac{D_a^{n-t}}{(n-t)!}\Big(a^{n-\nu}\Big).
\eeq
The last derivative is easily computed and we end up with
\beq
\wti{b}_{n-t+1}[F]\cong\sum_{l=1}^r\frac{\mu_l}{\tp_l}\sum_{s=1}^l\,(-\eps_1)^{s-1}\frac{\tp_{l-s+1}}{\De_1^{l-s+1/2}}\sum_{\nu=0}^t(-1)^{\nu}\,\si^{(1,\,s)}_{\nu-s}\,{n-\nu \choose t-\nu}\,a^{t-\nu}.
\eeq
We are now in position to compute the $b_k[F]$ which are given by
\beq
b_k[F]\cong (-1)^k\sum_{t=1}^k\,(-1)^t\,{ n-t \choose k-t}\,a^{k-t}\,\wti{b}_{n-t+1}.\eeq
Substituting into this formula the various pieces, exchanging the summation indices $\,\nu\leftrightarrow t$, and after 
some easy simplifications we get 
\beq\barr{l}\dst
b_k[F]\cong (-1)^k\sum_{l=1}^r\frac{\mu_l}{\tp_l}\,\sum_{s=1}^l(-\eps_1)^{s-1}\frac{\tp_{l-s+1}}{\De_1^{l-s+1/2}}
\sum_{\nu=0}^k\,(-1)^{\nu}\,
\si^{(1,\,s)}_{\nu-s}\,\frac{(n-\nu)!}{(n-k)!}\,a^{k-\nu}
\times
\\[4mm]\dst
\hspace{8cm}\times\,\sum_{t=\nu}^k\,\frac{(-1)^t}{(k-t)!\,(t-\nu)!}, \earr\eeq
and the sum over $t$ (using Lemma 1 in \cite{Va1})  simplifies to $(-1)^{\nu}\,\de_{k \nu}$. Adding the 
$\xi_i$ contribution we get (\ref{formQ1}). $\hfill\Box$

\nin{\bf Remarks:}
\brm
\item The formula obtained for the $b_k$ was proved under the restriction $r\in\{2,3,\ldots,n\}$. In fact it remains valid for $r=1$. In this case we have $\dst F(a)=\prod_{i=1}^n(a-a_i)$ leading back to the {\em simple} case. In the first piece of (\ref{formQ1}), substituting   $\mu_1\, \to\, \xi_1$ and noticing that 
$\tp_1=1$, leads to
\[ \xi_1\,\si^{(1,1)}_{k-1}\frac 1{\sqrt{\De_1}}=\xi_1\,\frac{\si^1_{k-1}}{\sqrt{\De_1}}\qq\Rightarrow\qq b_k=(-1)^k\sum_{i=1}^n\frac{\xi_i}{\sqrt{\De_i}}\,\si^i_{k-1},\]
giving indeed the correct result. 
\item It was shown in \cite{Va1} that the $b_k[F]$ should be a solution of the differential system
\beq\barr{ll}
k=0\quad & \hspace{4mm}b_1= -x\\[4mm]\dst 
1\leq k \leq n-1 \quad & b'_{k+1} = a\,b'_k+\frac 12\,b_k+(-1)^{k+1}\,\si_k\,x'\\[4mm]\dst 
k=n & \hspace{5mm}0  = a\,b'_n+\frac 12\,b_n+(-1)^{n+1}\,\si_n\,x'\earr\eeq
We have checked these relations, leaving the details of this computation to the interested reader.
\erm
Let us consider now the integral $S_2$.

\subsubsection{Determining ${\mathbf S_2}$}
The second integral is
\[ S_2=Q_2+y\,Q_1+\frac{y^2}{2}\,G\qq\qq\sharp[S_2]=\sharp[Q_2]=2n,\]
and as shown in \cite{Va1} we have
\beq\label{defQ2}
Q_2=\sum_{k=1}^n\,c_k[F]\,H^{n-k}\,P_y^{2k}\qq\qq c'_k[F]=-b_k[F]\,x'.
\eeq
Let us prove:

\begin{nth}\label{Q2} We have the relations
\beq
c_k[F]=\frac{(-1)^{k+1}}{2}\Big(c_k^{(\mu\,\mu)}+2\,c_k^{(\mu\,\xi)}+c_k^{(\xi\,\xi)}\Big),\eeq
where 
\beq\label{ck1}
c_k^{(\mu\,\mu)}=\sum_{l,m=1}^r\mu_l\,\mu_m\sum_{s=1}^l(-\eps_1)^{s-1}\,\si^{(1,s)}_{k-s}\,\frac {(m-1/2)\tp_{l-s+1}+(l-1/2)\tp_{m-s+1}}{(l+m-s)\,\De_1^{l+m-s}}\eeq
and
\beq\label{ck2}
c_k^{(\mu\,\xi)}=\sum_{l=1}^r\mu_l\sum_{i=r+1}^n\frac{\xi_i}{\sqrt{\De_i}}\sum_{s=1}^l(-\eps_1)^{s-1}\Big(\si^{(1,s)\,i}_{k-s}+a\,\si^{(1,s)\,i}_{k-s-1}\Big)\frac{\tp_{l-s+1}}{\De_1^{l-s+1/2}},\eeq
while the last piece is the same as for the {\rm simple} case
\beq\label{ck3}
c_k^{(\xi\,\xi)}=\sum_{i=r+1}^n\,\frac{\xi_i^2}{\De_i}\,\si^i_{k-1}+\sum_{i\neq j=r+1}^n\frac{\xi_i\,\xi_j}{\sqrt{\De_i\,\De_j}}\Big(\si^{ij}_{k-1}+a\,\si^{ij}_{k-2}\Big).
\eeq\end{nth}

\nin{\bf Proof:} Noticing that
\beq
-x'=\sum_{m=1}^r\,\eps_1(m-1/2)\,\frac{\mu_m\,\tp_m}{\De_1^{m+1/2}}+\sum_{j=r+1}^n\,\frac{\eps_j\,\xi_j}{2\De_j^{3/2}}\eeq
shows that
\beq
c_k^{(\mu\,\mu)}=-\sum_{l,m}\mu_l\,\mu_l\,(m-1/2)\sum_{s=1}^l(-\eps_1)^{s-1}\,
\si^{(1,s)}_{k-s}\,\tp_{l-s+1}\,
\int\frac{\eps_1\,da}{\De_1^{l+m-s+1}}.
\eeq
Integrating and symmetrizing 
$(l\,\leftrightarrow\,m)$ gives (\ref{ck1}).

In the next coefficient we have two pieces
\beq
c_k^{(\mu,\xi)}=-\sum_{l=1}^r\mu_l\,\sum_{i=r+1}^n\xi_i\left(\eps_1\si^i_{k-1}\,I^{1,\,i}_{\,l,0}+
\sum_{s=1}^l(-\eps_1)^{s-1}\,\si^{(1,s)}_{k-s}\,\tp_{l-s+1}\int\frac{\eps_i\,da}{2\De_i^{3/2}\,\De_1^{l-s+1/2}}\right)
\eeq
where
\beq
I^{1,\,i}_{\,l,\,0}=\int\frac{(l-1/2)\,da}{\De_1^{l+1/2}\,\sqrt{\De_i}}.
\eeq 
Integrating by parts the second piece, we are left with
\beq\barr{l}\dst 
c_k^{(\mu,\xi)}=\sum_{l,i}\mu_l\,\frac{\xi_i}{\sqrt{\De_i}}\sum_{s=1}^l(-\eps_1)^{s-1}\,\si^{(1,s)}_{k-s}\,\frac{\tp_{l+1-s}}{\De_1^{l-s-1/2}}\\[4mm]\dst 
\hspace{3cm}-\eps_1\sum_{l,i}\mu_l\xi_i\left(\si^i_{k-1}\,\tp_l\,I^{1,\,i}_{\, l,\,0}-\sum_{s=1}^l(-\eps_1)^{s-1}\,\si^{(1,s)}_{k-s}\,\tp_{l+1-s}\,I^{1,\,i}_{l+1-s,0}\right)\earr
\eeq
Using for the integrals $I^{1,\,i}_{\,l,\,0}$ and $I^{1,\,i}_{l-s+1,\,0}$ their explicit form given by Proposition \ref{intpart} gives  
\beq\barr{l}\dst 
c_k^{(\mu,\xi)}=\sum_{l,i}\mu_l\,\frac{\xi_i}{\sqrt{\De_i}}\sum_{s=1}^l(-\eps_1)^{s-1}\,\si^{(1,s)}_{k-s}\,\frac{\tp_{l+1-s}}{\De_1^{l-s-1/2}}\\[4mm]\dst 
\hspace{5cm}-\sum_{l,i}(\eps_1\eps_i)\mu_l\,\xi_i\sqrt{\De_i}\,\sum_{t=1}^l\frac{\tp_t}{\eta^{l+1-t}}\frac 1{\De_1^{t-1/2}}\,\Lambda^i_{k-1,l-t},\earr
\eeq
where
\beq
\Lambda^i_{k-1,l-t}=\si^i_{k-1}-\sum_{s=1}^{l-t+1}(-\eps_1\eta)^{s-1}\,\si^{(1,s)}_{k-s}.
\eeq
Using Proposition \ref{Bid} we have
\beq
\Lambda^i_{k-1,l-t}=(-\eps_1 \eta)^{l-t+1}\ \si^{(1,l-t+1)\,i}_{k-(l-t+2)},
\eeq
and changing the summation index $t\,\to\,s=l+1-t$ leads to 
\beq 
c_k^{(\mu,\xi)}= \sum_{l=1}^r\mu_l\,\sum_{i=r+1}^n\,
\frac{\xi_i}{\sqrt{\De_i}}\sum_{s=1}^l(-\eps_1)^{s-1}
\Big((a-a_i)\si^{(1,s)\,i}_{k-s-1}+\si^{(1,s)}_{k-s}\Big)\frac{\tp_{l-s+1}}{\De_1^{l-s+1/2}}.
\eeq
Using the identity (\ref{bid2}) we get (\ref{ck2}).

For $c_k^{(\xi,\xi)}$ the computation is the same as the one given in \cite{Va1}.$\hfill\Box$

\vspace{5mm}
\nin{\bf Remarks:} 
\brm
\item One can check again that for $r=1$ we recover the correct result for the {\em simple} case, which indeed deserves its name.
\item Specializing to the case $n=2$, the parametric solution of Novichkov equations is given this time by
\beq\left\{\barr{l}\dst 
{\cal H}(t)=x(a)=\frac{\mu_1}{\sqrt{a-a_1}}+\frac{\mu_2}{(a-a_1)^{3/2}}\\[4mm]\dst  
{\cal T}(t)=\frac{(2\mu_2-a_1\mu_1)}{\sqrt{a-a_1}}-\frac{a_1\,\mu_2}{(a-a_1)^{3/2}}\earr\right.  \qq h_t=\sqrt{a}, \eeq
where the constants $B_5$ and $B_6$ are now
\beq
B_5=4\frac{\mu_2}{a_1}(2\mu_2-a_1\,\mu_1)\qq\qq 
B_6=-4\frac{\mu_2}{a_1^2}(\mu_2-a_1\,\mu_1).\eeq
\erm

\subsection{The general case} 
Let us consider now the general case where 
\[F(a)=(a-a_1)^{r_1}\cdots(a-a_t)^{r_t}\,\wh{F} \qq \quad r=\sum_{\alf=1}^t\,r_{\alf}.\]
From Proposition \ref{Q1} we know, by linearity that we may take
\beq
x(a)=\sum_{\alf=1}^t\sum_{l=1}^{r_{\alf}}\,\mu_{\alf,\,l}\,\frac 1{\De_{\alf}^{l-1/2}}+\sum_{i=r+1}^n\frac{\xi_i}{\sqrt{\De_i}},
\eeq
and for the coefficients of $S_1$ we have
\beq 
b_k[F]=(-1)^k\left(\sum_{\alf=1}^t\sum_{l=1}^{r_{\alf}}\,\frac{\mu_{\alf,l}}{\tp_l}\sum_{s=1}^{l}(-\eps_{\alf})^{s-1}\,\si^{(
\alf,\,s)}_{k-s}\,\frac{\tp_{l-s+1}}{\De_{\alf}^{l-s+1/2}}+\sum_{i=3}^n\frac{\xi_i}{\sqrt{\De_i}}\si^i_{k-1}\right)
\eeq
in obvious notations. 

Considering now $S_2$  we have to compute
\beq
c_k[F](a)=-\int x'(a)\,b_k[F](a)\,da.\eeq
To this aim we will split again these coefficients
\beq
c_k=\frac{(-1)^{k+1}}{2}\Big(c_k^{(\mu\,\mu)}+2\,c_k^{(\mu\,\xi)}+c_k^{(\xi\,\xi)}\Big)
\eeq
and elementary computation gives
\beq\barr{l}\dst 
c_k^{(\mu\,\mu)}=2\sum_{\alf=1}^t\sum_{l,m=1}^{r_{\alf}}\mu_{l,\alf}\,\mu_{m,\alf}\sum_{s=1}^l
(-\eps_{\alf})^{s-1}\,\si^{(\alf,s)}_{k-s}\,\frac{l-1/2}{l+m-s}\,\frac{\tp_{m-s+1}}{\De_{\alf}^{l+m-s}}\\[4mm]\dst 
+2\sum_{\alf,\be=1}^t \sum_{l=1}^{r_{\alf}}\sum_{m=1}^{r_{\be}}\mu_{l,\alf}\,\mu_{m,\be}\sum_{s=1}^l
(-\eps_{\be})^{s-1}\,\si^{(\be,s)}_{k-s}\tp_{m-s+1}
\,(-\eps_{\alf})\,I^{\alf,\be}_{l,m-s},\earr
\eeq
where  $I^{\alf,\be}_{l,m-s}$ is given by Proposition \ref{intgen}.

No proof is required for the next coefficient which is
\beq
c_k^{(\mu\,\xi)}=\sum_{\alf=1}^t\sum_{l=1}^{r_{\alf}}\,\mu_{l,\alf}\sum_{i=r+1}^n\frac{\xi_i}{\sqrt{\De_i}}\sum_{s=1}^l(-\eps_{\alf})^{s-1}\Big(\si^{(\alf,s)\,i}_{k-s}+a\,\si^{(1,s)\,i}_{k-s-1}\Big)\frac{\tp_{l-s+1}}{\De_{\alf}^{l-s+1/2}}
\eeq
and follows from Proposition \ref{Q2} by linearity.

The coefficient $c_k^{(\xi,\xi)}$ remains the same as in Proposition \ref{Q2}.

\subsection{Globally defined examples}
We will give two examples, generalizing the previous ones given in \cite{Va1}.

The first one is a close cousin of the example given by Proposition 17 in \cite{Va1}. We will consider
\beq
F(a)=(a-1)^r\,\wh{F}(a)\qq\qq \wh{F}(a)=\prod_{i=r+1}^n(a-a_i)\qq 1\leq r\leq n.
\eeq
Let us prove:

\begin{nth} The hamiltonian $H$ and the integrals $(S_1,\,S_2)$ associated with the previous choice of $F$ are globally defined on $M\cong{\mb H}^2$.
\end{nth}

\nin{\bf Proof:} We will take \footnote{As usual if $r=n$ we take $\wh{F}=1$ and in $x$ the sum over $i$ should vanish.}
\beq
x(a)=\frac 1{2}\sum_{l=1}^r\,\frac{\mu_l}{(1-a)^{l-1/2}}-\sum_{i=r+1}^n\frac{\eps_i\xi_i(-\eps_i\,a_i)^{3/2}}{\sqrt{\eps_i(a-a_i)}}\qq\qq 0<a<1
\eeq
and the constraints
\beq
(a_i<0,\ \eps_i=+1) \quad \mbox{or} \quad (a_i>1,\ \eps_i=-1)\qq\qq \mu_l\geq 0 \qq\qq \xi_i\geq 0.\eeq 
The metric
\beq
g=\left(\frac{x'}{a}\right)^2\,da^2+\frac{dy^2}{a}
\eeq
under the coordinate change $\dst u=\sqrt{\frac a{1-a}}$ becomes
\beq
g=(1+u^2)\,\frac{\mu(u)^2\,du^2+dy^2}{u^2}\qq\qq u\in\,(0,+\nf)\qq y\in{\mb R}
\eeq
where
\[\mu(u)=\sum_{l=1}^r(l-1/2)\mu_l(1+u^2)^{l-1}+\sum_{i=r+1}^n\frac{\xi_i}{(1+\rho_i\,u^2)^{3/2}},\qq \rho_i=1-\frac 1{a_i}.\]
and
\[
\rho_i=1-\frac 1{a_i}\in\,(0,1)\cup(1,+\nf).
\]
Let us define
\beq
u\,\Om(u)=\int_0^u\mu(v)\,dv.\eeq
Thanks to Proposition \ref{Cid2} one obtains
\beq  
\Om(u)=\sum_{l=1}^r \,\frac{\mu_l}{2}\sum_{s=1}^l\,\tp_s\,(1+u^2)^{s-1}+\sum_{i=r+1}^n\frac{\xi_i}{\sqrt{1+\rho_i\,u^2}}.
\eeq
Let us define the new variable 
\[t=u\,\Om(u)\in\,(0,+\nf)\qq\longrightarrow\qq \frac{dt}{du}=\mu(u)>0.\]
We see that the inverse function $u(t)$ is increasing and $C^{\nf}([0,+\nf))$. The metric becomes
\beq
g=(1+u^2(t))\,\Om^2(u(t))\,\frac{dt^2+dy^2}{t^2},
\eeq
and since the $C^{\nf}([0,+\nf))$ conformal factor never vanishes, we conclude that $M\cong {\mb H}^2$.

We have seen that the integrals are
\beq
S_1=Q_1+y\,G\qq S_2=Q_2+y\,Q_1+\frac{y^2}{2}G\qq G=\sum_{k=0}^n\,A_k\,H^k\,P_y^{2(n-k)}
\eeq
and we know that $(H,\,P_y,\,G)$ are globally defined on $M$. 

Let us consider $Q_1$, defined in Proposition \ref{Q1}. Using the relation (\ref{formQ1}), we have
\beq\barr{l}\dst 
(-1)^k\,b_k=\sum_{l=1}^r\frac{\mu_l}{\tp_l}\sum_{s=1}^l\,\si^{(1,s)}_{k-s}\,\tp_{l-s+1}\,(1+u^2(t))^{l-s+1/2}\\[4mm]\dst \hspace{6cm}+\sum_{i=r+1}^na_i\,\xi_i\,\frac{\sqrt{1+u^2(t)}}{\sqrt{1+\rho_i\,u^2(t)}}\,\si^i_{k-1}.\earr
\eeq
All the functions of $t\in\,(0,+\nf)$ are indeed 
$C^{\nf}$ so that $Q_1$, hence $S_1$, is globally defined on $M$.

Using  the relations (\ref{ck1}), (\ref{ck2}) and (\ref{ck3}) the proof that $Q_2$, hence $S_2$, are globally defined on the manifold $M\cong{\mb H}^2$ is quite similar. $\hfill\Box$

The second example is given by the choice
\beq
F(a)=(a-a_1)^{r_1}(a-a_2)^{r_2}\,\wh{F}(a)\qq\qq r=r_1+r_2 \qq 1\leq r \leq n.
\eeq
Let us prove:

\begin{nth} The hamiltonian $H$ and the integrals $(S_1,\,S_2)$ associated with the previous choice of $F$ are globally defined on $M\cong{\mb R}^2$.
\end{nth}

\nin{\bf Proof:} We will start from \footnote{As usual if $r+s=n$ we take $\wh{F}=1$ and in $x$ the sum over $i$ should vanish.}
\beq
x(a)=-\sum_{l=1}^{r_1}\frac{\mu_l}{(a-a_1)^{l-1/2}}+\sum_{l=1}^{r_2}\frac{\nu_l}{(a_2-a)^{l-1/2}}-2\sum_{i=r+1}^n\,\frac{\eps_i\,\xi_i}{\sqrt{\De_i}}.
\eeq
with
\[ 0<a_1<a<a_2 \qq \mu_k> 0 \qq \nu_l> 0 \qq \xi_i> 0\qq a_i\not\in\,[a_1,a_2].\]
The derivative is
\beq
D_a\,x=\sum_{l=1}^{r_1}\frac{(l-1/2)\,\mu_l}{(a-a_1)^{l+1/2}}+\sum_{l=1}^{r_2}\frac{(l-1/2)\,\nu_l}{(a_2-a)^{l+1/2}}+\sum_{i=r+s+1}^n\,\frac{\xi_i}{\De_i^{3/2}},
\eeq
allows to define $\dst dt=\frac{dx}{\sqrt{a}}$. 
Using Proposition \ref{intpart} we 
get for the new coordinate $t(a)$ the explicit formula 
\beq\barr{l}\dst 
\frac{t(a)}{\sqrt{a}}=\sum_{l=1}^{r_1}\,\frac{\mu_l}{\tp_l}\,\sum_{s=1}^l\frac{\tp_s}{(-a_1)^{l-s+1}}\frac 1{(a-a_1)^{s-1/2}}\\[4mm]\dst
\hspace{2cm} +\sum_{l=1}^{r_2}\,\frac{\nu_l}{\tp_l}\sum_{s=1}^l\frac{\tp_s}{a_2^{l-s=1}}\frac 1{(a_2-a)^{s-1/2}}-\sum_{i=r+1}^n\frac{\xi_i}{a_i\,\sqrt{\De_i}}\earr
\eeq
which shows that $t(a)$ is an increasing $C^{\nf}$ bijection  from $a\in\,(a_1,a_2)\ \to\ t\in\,{\mb R}$ ensuring the existence of its inverse function $a(t)$ 
which maps ${\mb R}\ \to\ (a_1,a_2)$.

In the coordinates $(t,\,y)$ we get for the metric
\beq
g=\frac 1{a(t)}(dt^2+dy^2)\eeq
and since the $C^{\nf}$ conformal factor $a(t)$ never vanishes we conclude that the manifold is $M\cong {\mb R}^2$.

For the integral $S_1$ we have the coefficients
\beq\barr{l}\dst 
(-1)^k\,b_k[F]=-\sum_{l=1}^{r_1}\,\frac{\mu_l}{\tp_l}\sum_{s=1}^l(-1)^t\,\si^{(1,s)}_{k-s}\frac{\tp_{l-s+1}}{(a(t)-a_1)^{l-s+1/2}}\\[4mm]\dst 
\hspace{1.5cm}+\sum_{l=1}^{r_2}\,\frac{\nu_l}{\tp_l}\sum_{s'=1}^l\,\si^{(2,s')}_{k-s'}\frac{\tp_{l-s'+1}}{(a_2-a(t))^{l-s'+1/2}}
-2\sum_{i=r+1}^n\frac{\eps_i\xi_i}{\sqrt{\eps_i(a(t)-a_i)}}\,\si^i_{k-1}\earr
\eeq
and all the functions involved are indeed $C^{\nf}$ for $a(t)\in\,(a_1,a_2)$.

Since the coefficients appearing in $S_2$ are 
\beq
c_k[F]=-\int^a \,b_k[F]\,D_a\,x\,da, \qq a\in\,(a_1,a_2),
\eeq
the same arguments do apply. $\quad\Box$

Let us examine now the case where the integrals are of odd degree.

\section{Integrals of odd degree in the momenta}
Let us first recall the results obtained in \cite{Va1}.

\subsection{The simple case}
The hamiltonian and $F(a)$ remain unchanged
\beq
g=\left(\frac{x'}{a}\right)^2 da^2+\frac{dy^2}{a},\quad  H=\Pi^2+a\,P_y^2,\quad \Pi=\frac a{x'}\,P_a, \qq F(a)=a^n+\sum_{k=0}^{n-1}\,A_k\,a^k,
\eeq 
but this time
\beq
G=H^n\,P_y+\sum_{k=0}^{n-1}A_k\,H^k\,P_y^{2(n-k)+1} \qq\qq \sharp[G]=2n+1\qq n\geq 1.
\eeq
The extra integral is now
\beq
S_1=Q_1+yG \qq\qq Q_1=\sum_{k=0}^n\,\wti{b}_k(a)\Pi^{2k+1}P_y^{2(n-k)}\qq \sharp[S_1]=\sharp[Q_1]=2n+1.
\eeq
As shown in \cite{Va1} the $\wti{b}_k$ are given by
\beq
\forall k\in\{0,1,\ldots,n-1\}:\qq \wti{b}_k[F]=\sum_{s=1}^{k+1}\,\frac{D_a^{k+1-s} F}{(k+1-s)!}\,\frac{D_a^s x}{(1/2)_s};\qq \wti{b}_n=\nu_n\,\in\,{\mb R}\bs\{0\},
\eeq
where $\dst F(a)=a^n+\sum_{k=0}^{n-1}A_k\,a^k$ and $x(a)$ 
is a solution of the ODE
\beq
Op_n[F]\,x(a)=(n+1/2)\nu_n\,a.
\eeq
In the {\em simple} case we have
\beq
x(a)=\frac{\nu_n}{2}\,a+x^{=}(a),
\eeq
where $x^{=}(a)$ is given in (\ref{xpair}). It is again interesting to define new functions $b_k$ which allow to write
\beq
Q_1=\sum_{k=0}^n\,b_k[F]\,\Pi\,H^{n-k}\,P_y^{2k} \qq\qq  
b_k[F]=\sum_{s=0}^k{n-s \choose k-s}\,(-a)^{n-k}\,\wti{b}_{n-s}[F].
\eeq
In the {\em simple} case one has
\beq
b_k[\wh{F}]=(-1)^k\nu_n\,\si_k+b_k^{=}[\wh{F}],
\eeq
where $b_k^{=}[\wh{F}]$ is defined in (\ref{bcpair}). A second integral of degree $2n+1$ can be constructed:
\beq
S_2=Q_2+yQ_1+\frac{y^2}{2}G \qq Q_2=\sum_{k=0}^n\,c_k[F]\,H^{n-k}\,P_y^{2k+1}\qq \sharp[Q_2]=2n+1
\eeq
provided that
\beq
 D_ac_k[F]=-(D_a x)\,b_k[F].
\eeq
In the {\em simple} case we had
\beq 
c_k[\wh{F}]=\frac{(-1)^{k+1}}{2}\left(\nu_n^2a\si_k+2\nu_n\sum_{i=1}^n\frac{\xi_i}{\sqrt{\De_i}}(\si^i_k+a\si^i_{k-1})\right)+c_k^{=}[\wh{F}],
\eeq
where $c_k^{=}[\wh{F}]$  is defined in (\ref{bcpair}).

Let us consider now the case of a {\em multiple real root} for $F$.

\subsection{The case of a multiple real root}
Let it be supposed that a single real root of $F$, let us say $a_1$, has multiplicity $r$ while all others roots remain simple: 
\beq
r\in\,\{2,\ldots,n\}: \qq F(a)=(a-a_1)^r\,\wh{F}\qq\qq \wh{F}=\prod_{i=r+1}^n\,(a-a_i)
\eeq
To shorten matters we will give only the results, which follow easily from techniques developed either in \cite{Va1} or in Section 2.

Let us denote by $x^{=}(a)$ the solution obtained in Proposition 2. Similarly $b_k^{=}[F]$ and $c_k^{=}[F]$, the functions needed in the integrals $S_1$ and $S_2$, and given respectively in Proposition 3 and Proposition 4  of Section 2. Then we have
\beq
x(a)=x^{=}(a)+\frac{\nu_n}{2}\,a.
\eeq 
From this result we deduce the coefficients of $Q_1$:
\beq
b_k[F]=b^{=}_k[F]+(-1)^k\nu_n\,\si_k
\eeq
and the coefficients of $Q_2$:
\beq
c_k[F]=c_k^{=}[F]+\frac{(-1)^{k+1}}{2}\Big(c_k^{(\nu\nu)}+2c_k^{(\nu\xi)}+2c_k^{(\nu\mu)}\Big)
\eeq
with
\beq
c_k^{(\nu\nu)}=\nu_n^2\,a\,\si_k\qq\qq  
c_k^{\nu\xi}=\nu_n\sum_{i=r+1}^n\frac{\xi_i}{\sqrt{\De_i}}(\si^i_k+a\,\si^i_{k-1})   
\eeq
and
\beq
c_k^{(\nu\mu)}=\nu_n\,\sum_{l=1}^r\frac{\mu_l}{\tp_l}\left(\si_k\,\frac{\tp_l}{\De_1^{l-1/2}}+\sum_{s=1}^l(-\eps_1)^s\,\si^{(1,s)}_{k-s}\frac{\tp_{l-s+1}}{(2l-2s-1)\,\De_1^{l-s-1/2}}\right).
\eeq

\subsection{Globally defined metrics}
We will give a few examples generalizing and correcting Proposition 27 in \cite{Va1}.

We will take $F(a)=(a-a_1)^r$ with $1\leq r\leq n$. This choice allows to take
\beq
x(a)=\frac a2-c\sum_{l=1}^r\frac{\mu_l}{\De_1^{l-1/2}}\qq\quad \De_1=\eps_1(a-a_1)\eeq
and gives the metric
\beq
g=\frac 1a\left(\frac{dx^2}{a}+dy^2\right)\qq\qq y\in\,{\mb R}.\eeq
The coordinate change $u=\sqrt{a}$ allows to write
\beq
g=\frac 1{u^2}\Big(\mu(u)^2\,du^2+dy^2\Big)\qq u>0,\ y\in\,{\mb R}\eeq
where
\beq
\mu(u)=1+\eps_1c\sum_{l=1}^r\frac{(2l-1)\mu_l}{\De_1^{l+1/2}}.\eeq
Various choices of the parameters will lead to different examples of globally defined models. Let us begin with

\begin{nth} Provided that all $\mu_l$ are strictly positive, the choice
\beq
\mu(u)=1+\sum_{l=1}^r\frac{(2l-1)\mu_l}{(1-u^2)^{l+1/2}}
\eeq
leads to a globally defined SI model on 
$M\cong{\mb H}^2$.\end{nth}

\nin{\bf Proof:} We have taken $\eps_1=-1$ giving  
$\De_1=a_1-u^2$ which requires $a_1>0$ so we will set $a_1=1$ and we will have $u\in\,(0,1)$. This enforces $c=-1$ in order to avoid a zero of $\mu$, which would produce a curvature singularity, since we have for the sectional curvature
\beq
R=-\frac{\mu+u\mu'}{\mu^3}.\eeq

The function $\mu(u)$ is strictly increasing from $\dst \mu(0+)=1+\sum_{l=1}^r(2l-1)\mu_l$ up to $\mu(1-)=+\nf$ which allows to define the new coordinate
\beq
t=\int_0^u\,\mu(s)ds: \qq u\in (0,1)\ \to\ t\in\,(0,+\nf) \quad\Longrightarrow\quad t=u\,\Om(u)
\eeq
and Proposition \ref{Cid2} gives
\beq
\Om(u)=1+\sum_{l=1}^r\frac{\mu_l}{\tp_l}
\sum_{s=1}^l\frac{\tp_s}{(1-u^2)^{s-1/2}}.
\eeq
Let us notice that $t(u)$ is a $C^{\nf}$ bijection as well as its inverse function $u(t)$. It follows that the metric becomes
\beq
g=\Om^2(u(t))\,\frac{dt^2+dy^2}{t^2} \qq\qq \Big( t>0\quad y\in{\mb R}\Big),\eeq
and since $\Om$ never vanishes it follows that $g$ is conformal to the Poincar\'e metric for the hyperbolic plane and therefore $M\cong {\mb H}^2$.

The check for the integrals is trivial since the $b_k$ and the $c_k$ merely depend on functions of the form
\beq
\frac 1{(1-u^2)^{l-s-1/2}}\qq\qq l\in\{1,2,\ldots,r\} \qq  l-s\geq 1,
\eeq
which are all $C^{\nf}$ for $u\in\,(0,1)$.$\hfill\Box$

Let us now consider the case $\eps_1=1$.
\begin{nth} Provided that all $\mu_l$ are strictly positive, the choice
\beq
\mu(u)=1+\sum_{l=1}^r\frac{(2l-1)\mu_l}{(1+u^2)^{l+1/2}}
\eeq
leads to a globally defined SI model on 
$M\cong{\mb H}^2$.
\end{nth}

\nin{\bf Proof:} Here we have for the coordinates $(u>0\ y\in{\mb R})$. It follows that$\mu$ is strictly decreasing from $\dst \mu(0)=1+\sum_{l=1}^r(2l-1)\mu_l$ down to $\mu(+\nf)=1$. Defining
\beq
t=\int_0^u\,\mu(s)ds\qq\Longrightarrow\qq t=u\Om(u)
\eeq
and using Proposition \ref{Cid2} we get
\beq
\Om(u)=1+\sum_{l=1}^r\frac{\mu_l}{\tp_l}\sum_{s=1}^l\frac{\tp_s}{(1+u^2)^{s-1/2}}\eeq
which never vanishes. We then conclude as in the previous proof.$\hfill\Box$

\begin{nth}Provided that all $\mu_l$ are strictly positive and that $\dst\sum_{l=1}^r(2l-1)\mu_l<1$, the choice
\beq
\mu(u)=1-\sum_{l=1}^r\frac{(2l-1)\mu_l}{(1+u^2)^{l+1/2}}
\eeq
leads to a globally defined SI model on 
$M\cong{\mb H}^2$.\end{nth}

\nin{\bf Proof:} The difference with the previous case is that now $\mu$ is strictly increasing from $\mu(0)>0$ up to $\mu(+\nf)=1$. Defining the new coordinate $t$ as in the previous Proposition, we have, using Proposition \ref{Cid2}:  
\beq
\Om(u)=1-\sum_{l=1}^r\frac{\mu_l}{\tp_l}\sum_{s=1}^l\frac{\tp_s}{(1+u^2)^{s-1/2}}\eeq
which never vanishes for $u\geq 0$ due to \footnote{One has to use (\ref{idCls}) in Appendix C for $k=0$.}
\beq
\sum_{l=1}^r\frac{\mu_l}{\tp_l}\sum_{s=1}^l\frac{\tp_s}{(1+u^2)^{s-1/2}}\leq \sum_{l=1}^r\mu_l\sum_{s=1}^l\,\frac{\tp_s}{\tp_l}=\sum_{l=1}^r(2l-1)\mu_l<1.
\eeq
We then conclude as in the previous proof.$\hfill\Box$

The last case is
\begin{nth} Provided that all $\mu_l$ are strictly positive, the choice
\beq
\mu(u)=1+\sum_{l=1}^r\frac{(2l-1)\mu_l}{(u^2-1)^{l+1/2}}
\eeq
leads to a globally defined SI model on 
$M\cong{\mb R}^2$.\end{nth}

\nin{\bf Proof:} We have for the coordinates $(u>1\ y\in{\mb R})$ and $\mu$ is strictly decreasing from $\mu(1+)=+\nf$ down to $\mu(+\nf)=1$.

This time we have
\beq
\Om(u)=1-\sum_{s=1}^l\,(-1)^{l-s}\frac{\mu_l}{\tp_l}\frac {\tp_s}{(u^2-1)^{s-1/2}}:\qq u\in\,(1,+\nf)\ \to\ 
t\in\,{\mb R}.\eeq
The mapping $u\,\to\,t$ is a $C^{\nf}$ bijection, so the metric becomes
\beq
g=\frac 1{u^2(t)}(dt^2+dy^2)\qq\qq (t,y)\in\,{\mb R}^2\eeq
implying $M\cong{\mb R}^2$ since the conformal factor never vanishes.

We then conclude as in the previous proof.$\hfill\Box$

\nin{\bf Remarks:}
\brm
\item Our analysis, valid also for $r=1$, does correct Proposition 27 in \cite{Va1}. The analyses of the cases 
${\cal T}_{+-}$ and ${\cal T}_{--}$ were mistaken.
\item Here we have considered the case $F(a)=(a-a_1)^r$. It would be quite easy to generalize to $F(a)=(a-a_1)^r\,\wh{F}(a)$ where $\dst\wh{F}(a)=\prod_{i=r+1}^n\,(a-a_i)$ with appropriate restrictions on the parameters $a_i$.
\item The choice
\beq
F(a)=(a-a_1)^{r_1}(a-a_2)^{r_2}\wh{F}(a)\qq\quad  0<a_1<a<a_2,\eeq
generalizing the example given in the Proposition 30 in \cite{Va1} can be worked out. The details are left to the interested reader.
\erm

\section{Conclusion}
Many problems remain open for this class of SI models. Let us just quote a few:
\brm
\item The case where $F$ has complex conjugate roots, leading to
\[ F(a)=(a^2+1)^r\wh{F}(a)\qq\qq r\geq 1\]
is still unsolved.
\item A full classification of the globally defined cases remains undone.
\item The Zoll metrics are expected in the trigonometric case (see the Introduction) but this problem has been solved, up to now, only for cubic integrals in \cite{vds}. A generalization to all degrees would be of the greatest interest.
\item The approach followed in \cite{ms} for the cubic case by Matveev and Shevchishin and generalized later by Novichkov in  \cite{No} for the quartic case gives highly non-linear ODEs of first order which remain to be solved...In our solution we observed some kind of a bifurcation behaviour: given the parameters $(a_1,\,a_2)$ which appear in the parametric form of the solution, the very structure of this solution is somewhat different according to whether  $a_1=a_2$ or $a_1\neq a_2$. 
\item At the quantum level, any quantization will give   
$[\wh{H},\wh{P}_y]=0$, as explained in \cite{dv}, but 
finding some quantization scheme for which
\[
[\wh{H},\wh{S}_1]=[\wh{H},\wh{S}_2]=0
\]
would be very interesting but possibly difficult.
\erm

\begin{appendices}

\section{Appendix A}
This Appendix is devoted to the computation of
\beq
B_{p,k,l}[F]=\sum_{s=p}^k\,\frac{F^{(k-s)}}{(k-s)!}\frac{D_a^s\,x_l}{(1/2)_s}\qq\qq x_l=\frac 1{\De_1^{l-1/2}} \qq \De_1=\eps_1(a-a_1)\eeq
provided that
\beq\label{restrict} k\in{\mb N},\qq 0\leq p\leq k,\qq l\in{\mb N}\bs\{0\}.
\eeq
We will prove:

\begin{nth}\label{ap1} Under the restrictions  (\ref{restrict}) one has the relation
\beq\label{form1}
B_{p,k,l}[F]=\frac{(-\eps_1)^p}{\tp_l}\ \sum_{s=1}^l\frac{(p-1/2)_{l-s}}{(l-s)!}\ \De_1^{s-(l+p)+1/2}\ \frac{D_a^{k-p}}{(k-p)!}\left(\frac F{\De_1^s}\right).\eeq
\end{nth}

\nin{\bf Proof:} The change $s'=s-p$ gives
\beq
B_{p,k,l}[F]=\sum_{s=0}^{k-p}\frac{F^{(k-p-s)}}{(k-p-s)!}\ Y_{s+p,l}\qq\qq Y_{s+p,l}=\frac{D_a^{s+p}\,x_l}{(1/2)_{s+p}}.
\eeq
Computing the derivatives of $x_l$ gives
\beq
Y_{s+p,l}=(-\eps_1)^{s+p}\frac{{(l-1/2)_{s+p}}}{(1/2)_{s+p}}\,\De_1^{-l-(s+p)+1/2}.
\eeq
Combining the relation
\beq
\frac{{(l-1/2)_{s+p}}}{(1/2)_{s+p}}=\frac{(s+p+1/2)_{l-1}}{(1/2)_{l-1}}
\eeq
and
\beq
D_a^{l-1}(\De_1^{-s-p-1/2})=(-\eps_1)^{l-1}\,(s+p+1/2)_{l-1}\,\De_1^{-l-s-p+1/2}
\eeq
one gets eventually
\beq
Y_{s+p,l}=\frac{(-\eps_1)^{p+l-1}}{\tp_l}
\ \frac{D_a^{l-1}}{(l-1)!}\Big(\De_1^{-p+1/2}\ 
(-\eps_1)^s\,\De_1^{-s-1}\Big) \qq\quad {\cal P}_l=\frac{(1/2)_{l-1}}{(l-1)!}.
\eeq
Leibnitz formula gives
\beq
Y_{s+p,l}=\frac{(-\eps_1)^{p+l-1}}{\tp_l}
\ \sum_{t=0}^{l-1}\left(\frac{D_a^{l-1-t}}{(l-1-t)!}\,\De_1^{-p+1/2}\right)\ \frac{D_a^t}{t!}\Big((-\eps_1)^s\,\De_1^{-s-1}\Big)
\eeq
and the relation
\beq
\frac{D_a^t}{t!}\Big((-\eps_1)^s\,\De_1^{-s-1}\Big)=\frac{D_a^t}{t!}\Big(\frac{D_a^s}{s!}\,\De_1^{-1}\Big)=\frac{D_a^s}{s!}\Big(\frac{D_a^t}{t!}\,\De_1^{-1}\Big)=\frac{D_a^s}{s!}\Big((-\eps_1)^t\,\De_1^{-t-1}\Big)\eeq
gives eventually
\beq
Y_{s+p,l}=\frac{(-\eps_1)^{p+l-1}}{\tp_l}
\ \sum_{t=0}^{l-1}(-\eps_1)^t\,\left(\frac{D_a^{l-1-t}}{(l-1-t)!}\,\De_1^{-p+1/2}\right)\ \frac{D_a^s}{s!}\Big(\De_1^{-t-1}\Big).
\eeq
Plugging this into $B_{p,k,l}$ and using again Leibnitz formula we are led to
\beq
B_{p,k,l}[F]=\frac{(-\eps_1)^{p+l-1}}{\tp_l}
\ \sum_{t=0}^{l-1}(-\eps_1)^t\,\left(\frac{D_a^{l-1-t}}{(l-1-t)!}\,\De_1^{-p+1/2}\right)\ \frac{D_a^{k-p}}{(k-p)!}\Big(\frac F{\De_1^{t+1}}\Big),
\eeq
and the relation
\beq
D_a^{l-1-t}\ \De_1^{-p+1/2}=(-\eps_1)^{l-1-t}\,(p-1/2)_{l-1-t}\,\De_1^{t-l-p+3/2}\eeq
implies
\beq
B_{p,k,l}[F]=\frac{(-\eps_1)^p}{\tp_l}
\ \sum_{t=0}^{l-1}\,\frac{(p-1/2)_{l-1-t}}{(l-1-t)!}
\,\De_1^{t-l-p+3/2}\ \frac{D_a^{k-p}}{(k-p)!}\Big(\frac F{\De_1^{t+1}}\Big)
\eeq
and the change $s=t+1$ gives Proposition \ref{ap1}.
$\hfill\Box$

\section{Appendix B}
We will consider the case where the polynomial $F(\tau)$ has a real zero, say $a_1$, of order $r$.

\subsection{Definitions}
Let us begin with
\beq
2\leq r\leq n: \qq\qq F_n(\tau)=(\tau-a_1)^r\,\wh{F}_{n-r}(\tau)
\qq \wh{F}_{n-r}(\tau)=\prod_{i=r+1}^n\,(\tau-a_i)\eeq
The polynomial $\wh{F}_{n-r}(\tau)$ has only simple zeroes $a_i$ and if $r=n$ we take $\wh{F}_0(\tau)=1$. 
The subscripts, which we will omit in what follows,  indicate the degree w. r. t. $\tau$ of the various polynomials.

Let us define some new symmetric functions of the roots by
\beq\label{bdef1}
0\leq s \leq r: \qq \frac{F(\tau)}{(\tau-a_1)^s}=\sum_{k=0}^{n+s}\,(-1)^{k-s}\ \si^{(1,\,s)}_{k-s}\,\tau^{n-k}.
\eeq 
Since the degree in $\tau$ of the left hand side is $n-r$ we have a set of non-vanishing symmetric functions
\[ \si^{(1,\,s)}_0=1,\quad \si^{(1,\,s)}_1, \ \cdots,\quad  \si^{(1,\,s)}_{n-s}\]
while all others vanish
\[\si^{(1,\,s)}_{-s}=\cdots=\si^{(1,\,s)}_{-1}=0\qq\qq\si^{(1,\,s)}_{n+1-s}=\cdots=\si^{(1,\,s)}_n=0.\]

Let us simplify these notations for $s=0,1$ by setting
\[\si^{(1,\,0)}_k=\si_k\qq\qq\qq \si^{(1,\,1)}_{k-1}=\si^1_{k-1}\]
defined as usual by the relations
\beq
F(\tau)=\sum_{k=0}^{n}\,(-1)^{k}\ \si_{k}\,\tau^{n-k}\qq\qq
\frac{F(\tau)}{\tau-a_1}=\sum_{k=0}^{n+1}\,(-1)^{k-1}\ \si^1_{k-1}\,\tau^{n-k}\eeq

If $r<n$ we can still define
\beq\label{bdef2}
r+1\leq i \leq n: \qq\qq \frac{F(\tau)}{\tau-a_i}
=\sum_{k=0}^{n+1}\,(-1)^{k-1}\ \si^i_{k-1}\,\tau^{n-k}\eeq
and also
\beq\label{bdef3}
\frac{F(\tau)}{(\tau-a_1)^s(\tau-a_i)}=\sum_{k=0}^{n+s+1}\,(-1)^{k-s-1}\,\si^{(1,\,s)\ i}_{k-s-1}\,\tau^{n-k}.
\eeq

\subsection{Identities}
Multiplying relation (\ref{bdef1}) by $\tau-a_1$ gives 
\beq\label{bid1}
\si^{(1,\,s)}_{k-(s-1)}+a_1\,\si^{(1,\,s)}_{k-s}=\si^{(1,\,s-1)}_{k-(s-1)}\qq\qq k\in\,\{0,1,\ldots, n+s-1\}.\eeq
Multiplying relation (\ref{bdef2}) by $\tau-a_i$ gives
\beq\label{bid2}
k\in\,\{0,1,\ldots,n+s\}: \qq\qq \si^{(1,\,s)}_{k-s}=\si^{(1,\,s)\,i}_{k-s}+a_i\,\si^{(1,\,s)\ i}_{k-s-1}.
\eeq
Multiplying relation (\ref{bdef2})  by $\tau-a_1$ one gets
\beq\label{bid3}
k\in\,\{0,1,\ldots,n+s\}: \qq\qq \si^{(1,\,s-1)}_{k-(s-1)}=\si^{(1,\,s)\ i}_{k-(s-1)}+a_1\,\si^{(1,\,s)\ i}_{k-s}.
\eeq 
We will need also:

\begin{nth}\label{Bid} One has the relation
\beq
\Lambda^i_{k-1}\equiv\si^i_{k-1}-\sum_{s=0}^{l-1-t}(a_1-a_i)^s\ \si^{(1,\,s+1)}_{k-(s+1)}=(a_1-a_i)^{l-t}\ \si^{(1,\,l-t)\ i}_{k-(l-t)-1}.
\eeq\end{nth}

\nin{\bf Proof:} Let us compute the generating function
\beq
\sum_{k=0}^{n+1}\,(-1)^{k-1}\Lambda^i_{k-1}\,\tau^{n-k}.
\eeq
Its first piece is given by (\ref{bdef2}) and the second one, after interchanging the summations, is given by (\ref{bdef1}). The sum over $s$ is a trivial finite geometric series leading to
\beq
\frac{F(\tau)}{\tau-a_i}-\frac{F(\tau)}{\tau-a_i}\left[1-\left(\frac{a_i-a_1}{\tau-a_1}\right)^{l-t+1}\right]=(a_i-a_1)^{l-t+1}\,
\frac{\,\wh{F}(\tau)}{(\tau-a_1)^{l-t+1}(\tau-a_i)}
\eeq
Using the expansion given in (\ref{bdef3}) concludes the proof.$\hfill\Box$

\section{Appendix C}
Let us begin with some easy formulae:
\begin{nth}\label{Cid2} For $l\geq 1$ we have the relations
\beq\label{int2}
(2l-1)\int^u\,(1+t^2)^{l-1}\,dt=\frac 1{\tp_l}\,\sum_{s=1}^l\,\tp_s\,u\,(1+u^2)^{s-1},\quad u>0, \qq \tp_l=\frac{(1/2)_{l-1}}{(l-1)!},
\eeq
as well as
\beq\label{int3}
(2l-1)\int^u\frac{dt}{(t^2+\eps)^{l+1/2}}=\frac{\eps}{\tp_l}\,\sum_{s=1}^l\,\eps^{l-s}\,\tp_s\,\frac u{(u^2+\eps)^{s-1/2}}\qq 
\left\{\barr{l}\eps=+1,\ u\in\,{\mb R}\\[4mm]\eps=-1,\,u>1,\earr\right.
\eeq
and
\beq\label{int4}
(2l-1)\int^u\frac{dt}{(1-t^2)^{l+1/2}}=\frac 1{\tp_l}\,\sum_{s=1}^l\,\tp_s\,\frac u{(1-u^2)^{s-1/2}} \qq u\in\,(0,1).\eeq
\end{nth}

\nin{\bf Proof:} The derivative of the right hand side is, after some algebra:
\beq
(2l-1)\,(1+u^2)^{l-1}+\frac 1{\tp_l}\sum_{s=1}^{l-1}\Big[(2s-1)\tp_s-2s\tp_{s+1}\Big](1+u^2)^{s-1},
\eeq
and the sum vanishes. The proofs of (\ref{int3}) and (\ref{int4}) are similar.$\hfill\Box$

Let us observe that differentiating (\ref{int2}) and expanding in power series gives the identities
\beq\label{idCls}
\forall k \in\{0,1,\ldots,l-1\}: \qq \frac 1{\tp_l}\,\sum_{s=k+1}^l \,{s-1 \choose k} \tp_s=\frac{(2l-1)}{2k+1}{l-1 \choose k}.\eeq

We will need also:

\begin{nth} \label{intpart} For $a_{\alf}\neq a_{\be}$ we have
\beq\label{int0}
I^{\,\alf,\,\be}_{\ l,\,0}\equiv \int\frac{(l-1/2)\,da}{\De_{\alf}^{l+1/2}\,\De_{\be}^{1/2}}=\frac{\eps_{\be}\,\sqrt{\De_{\be}}
}{\tp_l}\sum_{s=1}^l\frac 1{\eta^s}\,\frac{\tp_{l-s+1}}{\De_{\alf}^{l-s+1/2}},\qq\quad \eta=\eps_{\alf}(a_{\be}-a_{\alf}).
\eeq
\end{nth}

\nin{\bf Proof:} Let us first observe that

\beq
\int\frac{da}{2\De_{\alf}^{3/2}\,\De_{\be}^{1/2}}=\frac{\eps_{\be}}{\eta}\,\sqrt{\frac{\De_{\be}}{\De_{\alf}}}.
\eeq
Defining ${\cal D}=\eps_{\alf}\,D_{a_{\alf}}$ we can write
\beq
I^{\,\alf,\,\be}_{\ l,\,0}=\frac 1{(1/2)_{l-1}}\,{\cal D}^{(l-1)}\int\frac{da}{2\De_{\alf}^{3/2}\,\De_{\be}^{1/2}}
=\frac{\eps_{\be}\sqrt{\De_{\be}}}{(1/2)_{l-1}}\,{\cal D}^{(l-1)}\left(\frac 1{\sqrt{\De_{\alf}}\,\eta}\right),
\eeq
 and using Leibnitz formula we get (\ref{int0}).
$\hfill\Box$

We need also its generalization:

\begin{nth}\label{intgen} One has, for $a_{\alf}\neq a_{\be}$, the relation
\beq\label{int1}
I^{\,\alf,\,\be}_{\ l,\,M}\equiv \int\frac{(l-1/2)\,da}{\De_{\alf}^{l+1/2}\,\De_{\be}^{M+1/2}}=-\frac{\eps_{\alf}}{\wti{\eta}}\,\frac{\De_{\be}^{1/2-M}}{\De_{\alf}^{l-1/2}}\,_2F_1\left(\barr{c} 1,\ -(l+M-1)\\ -l+3/2\earr;-\si\frac{\De_{\alf}}{\wti{\eta}}\right).
\eeq
valid for all $l \geq 1$, for all $M\geq 0$ and $\wti{\eta}=\eps_{\be}(a_{\alf}-a_{\be})\neq 0$. 
\end{nth}

\nin{\bf Proof:} The change of variable 
$u=\sqrt{\De_{\alf}}$ gives
\beq
I^{\,\alf,\,\be}_{\ l,\,M}=(2l-1)\,\eps_{\alf}\int\frac{du}{u^{2l}(\wti{\eta}+\si\,u^2)^{M+1/2}}\qq\qq \si=\eps_{\alf}\eps_{\be}.
\eeq
For $\eta>0$, expanding the denominator using the binomial theorem gives
\beq
I^{\,\alf,\,\be}_{\ l,\,M}=\frac{-\eps_{\alf}}{\eta^{M+1/2}}\sum_{s\geq 0}\frac{(M+1/2)_s}{s!}\left(-\frac{\si}{\wti{\eta}}\right)^s\frac{-2l+1}{-2l+2s+1}u^{-2l+2s+1}.
\eeq
Using the identity
\beq
\frac{-2l+1}{-2l+2s+1}=\frac{(-l+1/2)_s}{(-l+3/2)_s}
\eeq
we get for result
\beq
I^{\,\alf,\,\be}_{\ l,\,M}=\frac{-\eps_{\alf}}{\wti{\eta}^{M+1/2}\,u^{2l-1}}\,_2F_1\left(\barr{c} M+1/2,\ -l+1/2\\ -l+3/2\earr;-\frac{\si\,u^2}{\wti{\eta}}\right),
\eeq
which implies
\beq
I^{\,\alf,\,\be}_{\ l,\,M}=\frac{-\eps_{\alf}}{\wti{\eta}^{M+1/2}\,\De_{\alf}^{l-1/2}}\,_2F_1\left(\barr{c} M+1/2,\ -l+1/2\\ -l+3/2\earr;-\si\frac{\De_{\alf}}{\wti{\eta}}\right) .
\eeq
We have
\[ z\equiv -\si\frac{\De_{\alf}}{\wti{\eta}}=\frac{a-a_{\alf}}{a_{\be}-a_{\alf}} \qq\qq 1-z=\frac{\De_{\be}}{\wti{\eta}}\] and upon use of
\beq
_2F_1\left(\barr{c} a,\ b\\ c\earr;z\right)=(1-z)^{c-a-b}\,_2F_1\left(\barr{c} c-a,\ c-b\\ c\earr;z\right)
\eeq
we end up with the required result. By analytic continuation this result remains valid for any  
$\wti{\eta}\neq 0$ because the hypergeometric function is merely a polynomial of degree $l+M-1$ in its variable. 
$\hfill\Box$ 

The reader can check that for $M=0$ one indeed recovers relation (\ref{int0}).

\end{appendices}

\end{document}